\documentstyle[12pt]{book}
\begin{document}

\sloppy
\raggedbottom

\chapter*{THE LIMITS OF MATHEMATICS\\(Extended Abstract)}
\markright
{The Limits of Mathematics}
\addcontentsline{toc}{chapter}
{The limits of mathematics}
\section*{G. J. Chaitin\\
IBM Research Division\\
\it chaitin@watson.ibm.com}
\section*{}

In June 1994 I gave a five-day course on the limits of mathematics at
the University of Maine in Orono.  This course featured a new approach
to algorithmic information theory (AIT).  Four versions [1]--[4] of
the course notes for this course, each using a somewhat different
approach, are available.  To automatically obtain any one of them in
\LaTeX, for example chao-dyn/9407003, send e-mail to ``{\tt chao-dyn @
xyz.lanl.gov}'' with ``{\tt Subject: get 9407003}''.

AIT deals with program-size complexity.  I define the complexity
$H(X)$ of an object $X$ to be the size in bits of the smallest program
that can calculate $X.$  Up to now, to get elegant mathematical
properties for this complexity measure $H(X),$ I had to measure the size
of programs for an abstract universal Turing machine.  This gave the
right mathematical properties, but it was not a programming language
that anyone could actually use.  Now I have found a way to obtain the
correct program-size complexity measure of AIT by measuring the size
of programs in a series of powerful and easy to use programming
languages.  These programming languages are versions of LISP that I
have invented expressly for this purpose.  Which of these programming
languages one considers most natural is to a certain extent a matter
of personal taste.

What does AIT have to say concerning the limits of mathematics?  My
theory yields two fundamental information-theoretic incompleteness
theorems.  First of all, my theorem, originally going back to 1970,
that an $N$-bit formal axiomatic system cannot enable one to exhibit
any specific object $X$ with program-size complexity $H(X)$ greater
than $N+c$.  Secondly, my theorem, originally going back to 1986, that
an $N$-bit formal axiomatic system cannot enable one to determine more
than $N+c'$ scattered bits of the halting probability $\Omega$.  In
chao-dyn/9407003, $c = 2359$ bits and $c' = 7581$ bits.  In
chao-dyn/9407005, $c = 1127$ bits and $c' = 3689$ bits.  In
chao-dyn/9407006, $c = 994$ bits and $c' = 3192$ bits.  And in
chao-dyn/9407009, $c = 735$ bits and $c' = 2933$ bits.

I think I prefer the ``aggressive'' formulation in chao-dyn/9407009.
I can also make a case for the ``conservative'' formulation in
chao-dyn/9407003.  chao-dyn/9407005 and chao-dyn/9407006 are
the intermediate steps between chao-dyn/9407003 and
chao-dyn/9407009.

After the references we summarize chao-dyn/9407003 in a four-page
appendix.  The first page is a table summarizing the version of LISP
that is used.  The second page is an example of a program written in
this LISP.  The third page summarizes the definitions, and the fourth
page summarizes the results.

\section*{References}

\begin{itemize}
\item[{[1]}] G. J. Chaitin, ``The Limits of Mathematics,''
IBM Research Report RC 19646, e-print chao-dyn/9407003, July 1994, 270 pp.
\item[{[2]}] G. J. Chaitin, ``The Limits of Mathematics II,''
IBM Research Report RC 19660, e-print chao-dyn/9407005, July 1994, 255 pp.
\item[{[3]}] G. J. Chaitin, ``The Limits of Mathematics III,''
IBM Research Report RC 19663, e-print chao-dyn/9407006, July 1994, 239 pp.
\item[{[4]}] G. J. Chaitin, ``The Limits of Mathematics IV,''
IBM Research Report RC 19671, e-print chao-dyn/9407009, July 1994, 231 pp.
\end{itemize}

\newpage
\begin{center}
\begin{tabular}{||c|l|l|l||}   \hline\hline
' & quote     & 1 arg  & '(abc) $\longrightarrow$ (abc)             \\ \hline
+ & head      & 1 arg  & +'(abc) $\longrightarrow$ a                \\
  &           &        & +a $\longrightarrow$ a                     \\ \hline
--& tail      & 1 arg  & --'(abc) $\longrightarrow$ (bc)            \\
  &           &        & --a $\longrightarrow$ a                    \\ \hline
* & join      & 2 args & *a'(bc) $\longrightarrow$ (abc)            \\
  &           &        & *ab $\longrightarrow$ a                    \\ \hline
. & atom      & 1 arg  & .a $\longrightarrow$ 1                     \\
  &           &        & .'(a) $\longrightarrow$ 0                  \\ \hline
= & equal     & 2 args & =aa $\longrightarrow$ 1                    \\
  &           &        & =ab $\longrightarrow$ 0                    \\ \hline
/ & if        & 3 args & /0ab $\longrightarrow$ b                   \\
  &           &        & /xab $\longrightarrow$ a                   \\ \hline
\& & function & 2 args & ('\&(xy)y ab) $\longrightarrow$ b          \\ \hline
, & display   & 1 arg  & ,x $\longrightarrow$ x and displays x      \\ \hline
! & eval      & 1 arg  & !e $\longrightarrow$ evaluate e            \\ \hline
? & try       & 3 args & ?teb $\longrightarrow$ evaluate e time t with bits b
\\
  &           &        & ?teb $\longrightarrow$
$
  \left(
     \begin{array}{c}
        \mbox{!} \\
        \mbox{?} \\
        \mbox{(value)}
     \end{array}
     \begin{array}{l}
        \mbox{captured} \\
        \mbox{displays\ldots}
     \end{array}
  \right)
$
                                                                    \\ \hline
@ & read bit  & 0 args & @ $\longrightarrow$ 0 or 1                 \\ \hline
\% & read exp & 0 args & \% $\longrightarrow$ any s-expression      \\ \hline
\# & bits for & 1 arg  & \#x $\longrightarrow$ bit string for x     \\ \hline
\verb|^| & append    & 2 args & \verb|^|'(ab)'(cd) $\longrightarrow$ (abcd)
  \\ \hline
\verb|~| & show      & 1 arg  & \verb|~|x $\longrightarrow$ x and may show x
  \\ \hline
: & let       & 3 args & :xv e    $\longrightarrow$ ('\&(x)e v)     \\
  &           &        & :(fx)d e $\longrightarrow$ ('\&(f)e '\&(x)d) \\ \hline
\& & define   & 2 args & \&xv    $\longrightarrow$ x is v           \\
  &           &        & \&(fx)d $\longrightarrow$ f is \&(x)d      \\ \hline
" & literally & 1 arg  & "+ $\longrightarrow$ +                     \\ \hline
\{\} & unary  &        & \{3\} $\longrightarrow$ (111)              \\ \hline
[ ]& comment  &        & [ignored]                   \\ \hline
( )& empty    &        &                             \\ \hline
0  & false    &        &                             \\ \hline
1  & true     &        &                             \\ \hline\hline
\end{tabular}
\end{center}

\newpage
\begin{verbatim}
lisp.c

LISP Interpreter Run

[[[(Fx) = flatten x by removing all interior parentheses]]]
[Define F of x as follows: if x is empty then return empty, if
 x is an atom then join x to the empty list, otherwise split
 x into its head and tail, flatten each, and append the results.]
& (Fx) /=x()() /.x*x() ^(F+x)(F-x)

F:          (&(x)(/(=x())()(/(.x)(*x())(^(F(+x))(F(-x))))))

(F,F) [use F to flatten itself]

expression  (F(,F))
display     (&(x)(/(=x())()(/(.x)(*x())(^(F(+x))(F(-x))))))
value       (&x/=x/.x*x^F+xF-x)

[[[(Gx) = size of x in unary]]]
[Let G of x be [if x is empty, then unary two, if x is an atom,
 then unary one, otherwise split x into its head and tail,
 size each, and add the results] in ...]
: (Gx) /=x()'{2} /.x'{1} ^(G+x)(G-x)
[Let G of x be [...] in:]
(G,G) [apply G to itself]

expression  (('(&(G)(G(,G))))('(&(x)(/(=x())('(11))(/(.x)('(1)
            )(^(G(+x))(G(-x))))))))
display     (&(x)(/(=x())('(11))(/(.x)('(1))(^(G(+x))(G(-x))))
            ))
value       (1111111111111111111111111111111111111111111111111
            111)

End of LISP Run

Elapsed time is 0 seconds.
\end{verbatim}

\newpage
{\Huge\bf DEFINITIONS}

\begin{itemize}
\item
An S-expression $x$ is elegant if no smaller S-expression has the same
output.  (Here ``output'' may be either its value or what it
displays.)
\item
Let $x$ be an S-expression.  The LISP complexity $H_L(x)$ of $x$ is
the size in characters $|p|$ of the smallest S-expression $p$ whose
value is $x$.
\item
Let $X$ be an infinite set of S-expressions.  The LISP complexity
$H_L(X)$ of the infinite set $X$ is the size in characters $|p|$ of
the smallest S-expression $p$ that displays the elements of $X$.
\item
\begin{verbatim}
[U(p) = output of universal machine U]
[       given binary program p.      ]
& (Up) ++?0'!%p
\end{verbatim}
\item
Let $x$ be an S-expression.  The complexity $H(x)$ of $x$ is the
smallest possible value of 7 times (the size in characters $|p|$ of an
S-expression $p$ whose value is $x$ if it is given the binary data
$d$) plus (the size in bits $|d|$ of the binary data $d$ given to
$p$).
\item
Equivalently $H(x) \equiv H_U(x)$ is the size in bits $|p|$ of the
smallest bit string $p$ such that $U(p) = x.$
\item
The halting probability $\Omega$ of $U$ is the limit as $t \rightarrow
\infty$ of (the number of $t$-bit programs $p$ such that $U(p)$ halts
within time $t$) divided by $2^t.$
\item
Let $X$ be an infinite set of S-expressions.  The complexity $H(X)$ of
the infinite set $X$ is the smallest possible value of 7 times (the
size in characters $|p|$ of an S-expression $p$ that displays the
elements of $X$ if it is given the binary data $d$) plus (the size in
bits $|d|$ of the binary data $d$ given to $p$).
\item
Equivalently $H(X) \equiv H_U(X)$ is the size in bits $|p|$ of the
smallest bit string $p$ such that $X = \lim_{t\rightarrow\infty}$ {\tt
-?t'!\%p.}
\end{itemize}

\newpage
{\Huge\bf RESULTS}

\begin{itemize}
\item
Lowcase variables $x, y, n$ are individual S-expressions.
\item[]
Uppercase variables $X, Y, T$ are infinite sets of S-expressions.
\item
$H_L (x,y) \le H_L(x) + H_L(y) + 8.$
\item
If $ x \in T \Longrightarrow x$ is elegant, then
\item[]
$x \in T \Longrightarrow |x| \le H_L(T) + 378.$
\item
If $(x,n) \in T \Longrightarrow H_L(x) \ge n,$ then
\item[]
$(x,n) \in T \Longrightarrow n \le H_L(T) + 381.$
\item
$H(x,y) \le H(x) + H(y) + 140.$
\item
Let $x$ be a string of $|x|$ bits.
\item[]
$H(x) \le 2|x| + 469,$ and $H(x) \le |x| + H(|x|) + 1148.$
\item
Let $\Omega_n$ be the first $n$ bits of $\Omega.$
\item[]
$H(\Omega_n) > n - 4431.$
\item
$H(X \cap Y) \le H(X) + H(Y) + 4193.$
\item
$H(X \cup Y) \le H(X) + H(Y) + 4193.$
\item
If $(x,n) \in T \Longrightarrow H(x) \ge n,$ then
\item[]
$(x,n) \in T \Longrightarrow n \le H(T) + 2359.$
\item
$T$ cannot determine more than $H(T) + 7581$ bits of $\Omega.$
\end{itemize}

\end{document}